\documentclass[prd,singlecolumn,amsmath,amssymb,nobibnotes,nofootinbib,superscriptaddress,preprintnumbers,12pt]{revtex4}
\usepackage{graphicx}

\newcommand{\be}{\begin{equation}}
\newcommand{\ee}{\end{equation}}
\newcommand{\bea}{\begin{eqnarray}}
\newcommand{\eea}{\end{eqnarray}}

\begin{document}

\title{Eternal Inflation With Non-Inflationary Pocket Universes}

\author{Jean-Luc Lehners}
\affiliation{Max-Planck-Institute for Gravitational Physics (Albert-Einstein-Institute), D-14476 Potsdam/Golm, Germany}

\begin{abstract}
Eternal inflation produces pocket universes with all physically allowed vacua and histories. Some of these pocket universes might contain a phase of slow-roll inflation, some might undergo cycles of cosmological evolution and some might look like the galilean genesis or other ``emergent'' universe scenarios. Which one of these types of universe we are most likely to inhabit depends on the measure we choose in order to regulate the infinities inherent in eternal inflation. We show that the currently leading measure proposals, namely the global light-cone cut-off and its local counterpart, the causal diamond measure, as well as closely related proposals, all predict that we should live in a pocket universe that starts out with a {\it small} Hubble rate, thus favoring emergent and cyclic models. Pocket universes which undergo cycles are further preferred, because they produce habitable conditions repeatedly inside each pocket.
\end{abstract}

\maketitle

If eternal inflation occurs, it generates pocket universes with all possible vacua. Moreover, since each type of universe is itself produced an infinite number of times, all possible histories within these pocket universes are physically realized \cite{Guth:2007ng,Aguirre:2007gy,Freivogel:2011eg}. This implies that we should not restrict ourselves to considering only pocket universes in which slow-roll inflation occurs, but in fact {\it all} physically allowed cosmological models will be realized an infinite number of times. 

Eternal inflation and ordinary inflation are, notwithstanding the similarity in terminology, rather separate concepts. Ordinary inflation is designed to dynamically generate physical conditions that resemble those we know to have been present some $13.7$ billion years ago in our region of the universe \cite{Guth:1980zm,Linde:1981mu,Albrecht:1982wi}. For this idea to work, it is essential that the phase of inflation occur for long enough to render the universe sufficiently flat, and inflation must occur at the right energy scale in order to produce density perturbations with an amplitude that is in agreement with the size of the temperature fluctuations that have been measured in the cosmic microwave background radiation. Modeling such a phase with a scalar field and a potential, the requirement is that the potential must contain a region that is flat over an extended region, and at a height of some $(10^{15} \, \rm{GeV})^4$ or so. 

By contrast, (false-vacuum) eternal inflation needs far less specific conditions in order to operate successfully: it is sufficient for there to exist a single sufficiently long-lived metastable vacuum with a positive cosmological constant. By sufficiently long-lived, we mean that the decay rate $\Gamma$ of this vacuum must be smaller than its expansion rate, $\Gamma < H^4.$ Then, starting with a spatially finite universe, regardless of what field configuration one starts with there is a non-zero probability for the spacetime to transition to a metastable vacuum with a positive cosmological constant, which means that eternal inflation {\it will} occur \cite{Brown:2011ry}. Once underway, a horizon-sized region can form a new pocket universe via quantum tunneling (mediated by a Coleman - De Luccia instanton \cite{Coleman:1980aw}). What physics takes place in this newly formed bubble universe depends on the properties of the potential on the other side of the potential barrier. If the potential is sufficiently flat after tunneling, the new pocket universe will start out with a period of inflation. If, on the other hand, the potential is that of a cyclic universe, and if cosmic bounces can occur (which remains an open question at present), then the new pocket universe will undergo cycles of evolution. And a horizon-sized region within the new pocket universe can itself tunnel back to the original metastable vacuum, or tunnel to a new vacuum of the theory under consideration. Any such tunneling processes that have a non-zero probability of occurring will of course occur in due time. In this way, the whole landscape gets populated. 

The fact that eternal inflation will produce all possible cosmological models begs the question as to which type of cosmology is more likely. This question is far from trivial, and sensitively depends on the measure that we choose in order to regulate the infinite number of universes that are generated. Historically, the first measure to be investigated in detail was the global proper time cut-off \cite{Linde:1993xx,Vilenkin:1994ua}. This prescription amounts to choosing a fixed proper time (measured along timelike geodesics starting from the initial spatial hypersurface that one assumes) and counting all pocket universes that have nucleated prior to this time. The relative number of different types of pockets is then taken to indicate their relative probability. At the end of the calculation, one takes the chosen time to future infinity, and the corresponding probabilities are used to make predictions (in the $t \rightarrow \infty$ limit, these relative probabilities become time-independent). By many people, this proposal made/makes the most intuitive sense. This measure is equivalent to weighting universes in proportion to their physical volume, and thus this measure predicts that the fastest growing pocket universes will come to dominate all probabilities. In this way, inflationary vacua with a high scale of inflation come out as preferred. However, precisely because this measure hugely rewards the fastest growing vacua, it turns out that it predicts that it is even more likely to be in a universe that nucleated only a fraction of a second ago from the fastest-expanding vacuum, and produced us via a quantum fluctuation rather than via ordinary evolution (note that the horizon in the fastest-growing vacuum is too small to accommodate a complex object of the size of a brain, otherwise it would have been even likelier to fluctuate directly out of the fastest-expanding vacuum) \cite{Bousso:2007nd}. Such non-sensical predictions have been termed the ``youngness paradox'' and have led to the eventual abandonment of this measure. 

Any measure that wants to avoid the youngness paradox is required to reward pocket universes less (or not at all) for their volume. This simple observation immediately implies that pocket universes with non-inflationary cosmologies should be taken into account too when calculating probabilities, as it is then far less clear why they should be unimportant. In fact, as we will argue here, it is the reverse that happens, and the currently leading measure proposals (which do avoid the youngness paradox) turn out to favor precisely those pocket universes that start out with a very small expansion rate. The basic reason for this is simple, but to state it we must first describe the measures which we work with. 

Principally, we will consider the lightcone time cut-off measure \cite{Garriga:2005av}. This measure works essentially in the same way as the proper time cut-off, but with a differently defined time variable. For a given event in spacetime, one can construct its future lightcone. This lightcone will intersect the boundary at future infinity. Using timelike geodesics, one can follow this intersected region back in time to the initial hypersurface with which one started. On this hypersurface, the region one has obtained in this way occupies a volume $v.$ Lightcone time is then defined as 
\be
t_{\rm{lc}} \equiv - \frac{1}{3}\ln v.
\ee
As shown in \cite{Garriga:1997ef,Garriga:2005av}, the lightcone cut-off measure leads to an attractor regime in the future, which implies that the most likely vacuum to be in is the longest-lived vacuum with a positive cosmological constant. This vacuum is often called the ``master vacuum'', and it dominates all probabilities that are calculated using this measure. 

Another measure proposal of interest, and which at first sight seems to be entirely unrelated, is the causal diamond measure \cite{Bousso:2006ev}. The prescription for calculating probabilities using this measure is to follow a single worldline, and to count the number of times that the worldline enters different vacua. This corresponds to counting all nucleation events that occur within the past lightcone of the point where the worldline reaches the future boundary of spacetime. Relative numbers of events then once more correspond to relative probabilities. As discovered by Bousso \cite{Bousso:2009dm} (see also \cite{Bousso:2009mw}), the causal diamond measure gives equivalent predictions to the lightcone cut-off measure if the worldline under consideration starts out in the master vacuum. Moreover, there are two closely related measure proposals, the scale-factor time cut-off \cite{DeSimone:2008bq} and the fat geodesic measure \cite{Bousso:2008hz}, which give essentially equivalent probabilities. The prescription for the fat geodesic measure is that once more one considers a worldline that starts out in the master vacuum, but then one counts only events that occur within a given physical distance from the worldline. Via a similar global-local duality as for the lightcone/causal diamond pair, probabilities then coincide with those calculated using the scale-factor time cut-off.

The bottom line of this discussion is that the currently leading measure proposals all lead to the same conclusion, namely that the most likely place to be in the multiverse is in the master vacuum. Should the master vacuum happen to be uninhabitable, then the most likely place to find oneself in becomes a pocket universe which is habitable and which can arise via the fastest tunneling sequence starting from the master vacuum. As discussed by Douglas \cite{Douglas:2012bu}, it is in fact very likely that the master vacuum itself is unsuitable for complex structures to form. Indeed, in order to be as long-lived as possible, the master vacuum should be as close as possible to being supersymmetric. Supersymmetric vacua are stable, and therefore infinitely long-lived. However, they cannot accommodate a positive cosmological constant. Thus, in the master vacuum, supersymmetry must be broken. But it should be broken as slightly as possible, in order for the vacuum to be as long-lived as possible. This implies that the cosmological constant in the master vacuum is likely to be very small (see also \cite{Banks:2000fe,Aguirre:2006ap}). One cannot directly estimate how small, because nothing else is known about the master vacuum. However, it seems reasonable to assume that it will be as small, or even smaller, than the present-day cosmological constant in our universe. The fact that supersymmetry should be only very slightly broken also implies that the master vacuum itself is likely unsuitable for life. Supersymmetry is a symmetry between bosons and fermions (and thus between forces and matter) and therefore seems unlikely to allow for complex structures to form. Certainly, it does not allow for life as we know it.

\begin{figure}[t]
\begin{center}
\hspace{-6cm}
\includegraphics[width=0.85\textwidth]{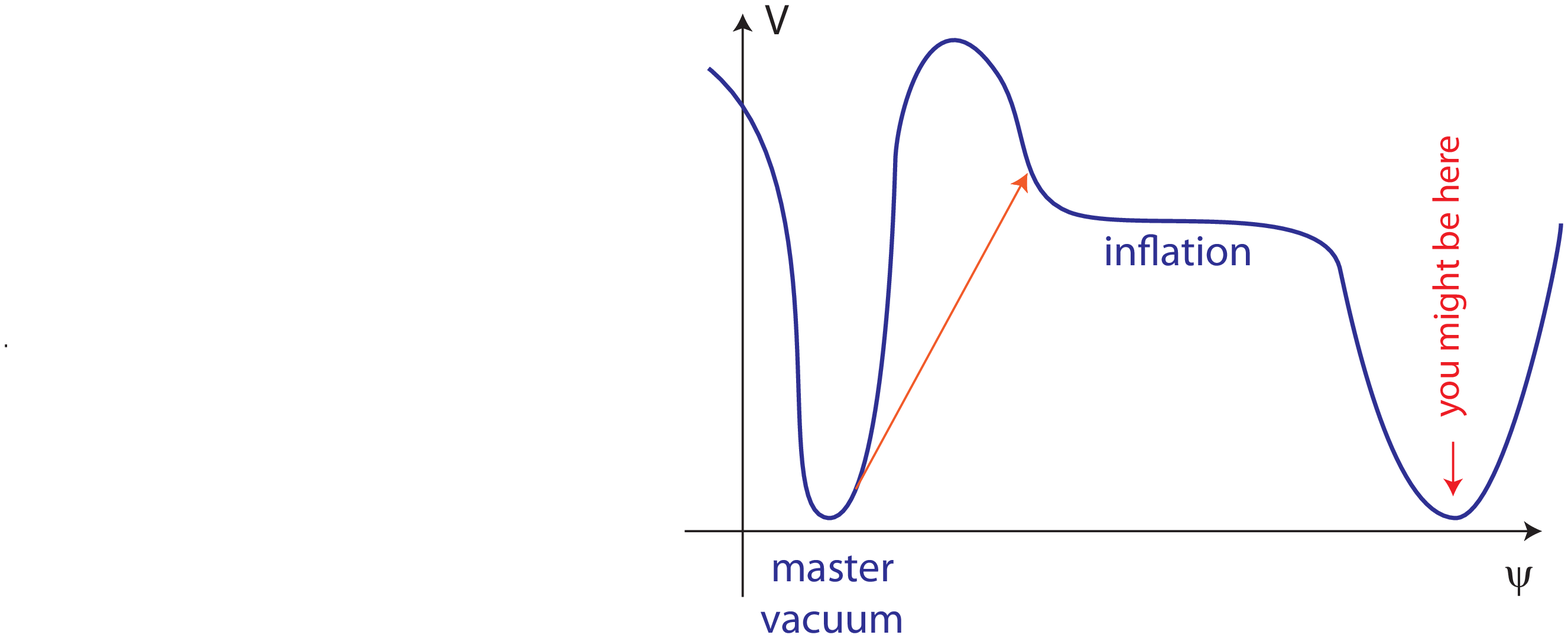}
\caption{\label{Figure1} {\small From the master vacuum, tunneling up to an inflationary pocket universe along the $\psi$ direction is possible, but the corresponding rate is small. 
}}
\end{center}
\end{figure}

\begin{figure}[h]
\begin{center}
\hspace{-6cm}
\includegraphics[width=0.85\textwidth]{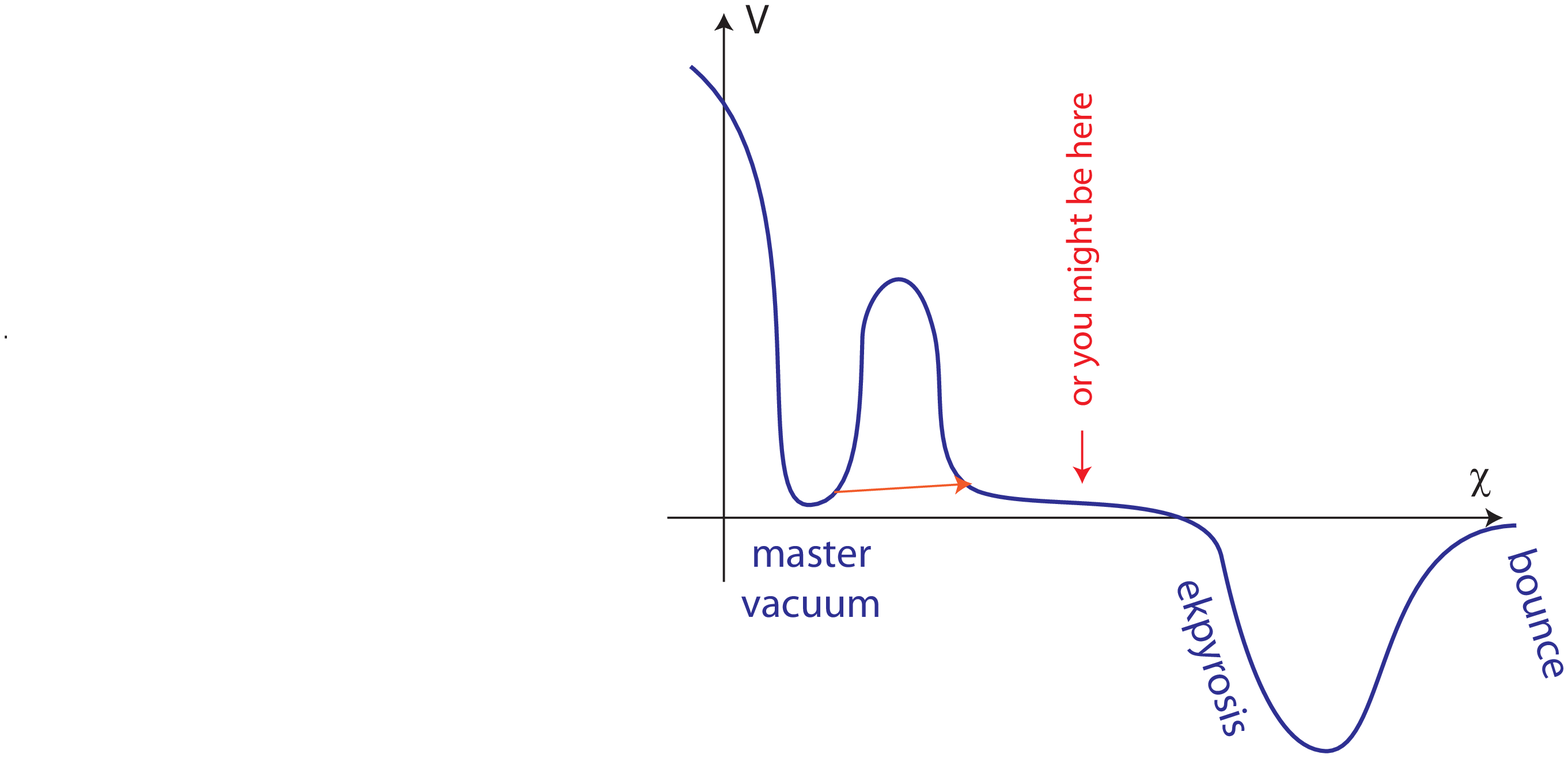}
\caption{\label{Figure2} {\small Tunneling from the same master vacuum along a different direction $\chi$ to a cyclic pocket universe can occur with a much faster rate.
}}
\end{center}
\end{figure}

Thus it seems clear that we should preferentially live in a pocket universe that arises via a fast tunneling process from the master vacuum. Producing a pocket universe inside of which inflation occurs then requires a large {\it up-tunneling} from the master vacuum, either directly, as depicted in Fig. \ref{Figure1}, or via some intermediate vacuum. Susskind has used this observation of the required proximity (in the landscape) to the master vacuum to argue that the scale of inflation might consequently be expected to be low \cite{Susskind:2012xf}. However, it seems that the logical conclusion of this argument leads to a different expectation, namely that universes which start out with a small expansion rate, and which can be reached via an approximately {\it equal-height} or even a {\it down-tunneling} process are vastly preferred -- see Fig. \ref{Figure2}. Indeed, there is no reason for the barrier between the master vacuum and the small-Hubble rate/small density vacua to be as large as between the master vacuum and the inflationary ones, considering that the difference in energy density between the small-$H$ vacua and the inflationary ones amounts to about $100$ orders of magnitude! This implies that pocket universes realizing galilean genesis \cite{Creminelli:2010ba} or other emergent universe scenarios \cite{Ellis:2002we,Joyce:2011ta} (in which the energy density starts out small and grows large), or those realizing the cyclic universe \cite{Steinhardt:2001st,Lehners:2008vx}, are vastly preferred over the usually exclusively considered inflationary pockets. Furthermore, because cyclic universes can produce habitable conditions repeatedly inside each pocket, they are further preferred over non-cyclic models. 

Now let us present the calculation which underlies these claims. It is very simple, and builds directly on the results of \cite{Johnson:2011aa}. We use the causal diamond measure, with the initial condition that the entire universe starts in the master vacuum, to perform the calculation. We consider seven different phases, allowing us to treat cyclic and non-cyclic universes in a unified manner. The master vacuum itself is denoted by $M.$ It can up-tunnel to an inflationary vacuum $I,$ with rate $\Gamma_{IM}.$ In general, the symbol $\Gamma_{ji}$ stands for the rate of transitioning from the phase $i$ to the phase $j.$ For simplicity, we amalgamate the inflationary phase, the subsequent radiation and matter phases and the eventual dark energy dominated phase. If one is interested in more specific questions, such as for example the Boltzmann brain problem, one needs to separate out these phases, as was done for example in \cite{Johnson:2011aa}. For our purposes, this is however not necessary. We also assume that the inflationary pocket can tunnel back to $M$ and that it can decay to a sink $S.$ Again, for simplicity we denote all terminal vacua collectively by $S.$ The non-inflationary pockets are separated out into four phases: the first, denoted $D_1$ is the small-$H$ dark energy-like phase that occurs right after bubble nucleation. We assume that this phase lasts for a time $t_{D_1},$ after which it transitions with probability $p_{\rm{first}}$ to a hot big bang phase $H_1.$ We do not want to assume that this automatically happens everywhere, hence the inclusion of the factor $p_{\rm{first}}.$ After a time $t_{H_1}$ this phase goes over to a new dark energy dominated phase $D$ which lasts for a time $t_D.$ To allow for cycling, we assume that after a time $t_D$ this phase can go over to a new hot phase $H$ with probability $p_{\rm{cycle}}$ and then back to $D$ again after a time $t_H.$ From the dark energy phases $D_1$ and $D$ we allow for down-tunneling to the sink $S$ as well as tunneling back to the master vacuum $M.$ Perhaps we should highlight that for emergent scenarios, the habitable phase is $H_1,$ while inside cyclic pockets ordinary observers can reside in $H_1$ as well as in any repeating occurrence of $H.$ Denoting by $f_i$ the fractional co-moving volumes occupied by vacuum $i$ leads to the rate equations
\bea
\dot{f}_M &=& -\Gamma_{D_1M}f_M - \Gamma_{IM} f_M + \Gamma_{MI}f_I +\Gamma_{MD_1} f_{D_1}+ \Gamma_{MD} f_{D}  \label{eq:req-infcyc-BB-F} \\
\dot{f}_I &=& \Gamma_{IM} f_M - (\Gamma_{MI}+\Gamma_{SI}) f_I \label{eq:req-infcyc-BB-I} \\
\dot{f}_{D_1} &=& \Gamma_{D_1M} f_M - \Gamma_{MD_1} f_{D_1} -\frac{1}{t_{D_1}} f_{D_1} \label{eq:req-infcyc-BB-D1} \\
\dot{f}_{H_1} &=& \frac{p_{\rm{first}}}{t_{D_1}} f_{D_1} - \frac{1}{t_{H_1}} f_{H_1} \label{eq:req-infcyc-BB-C1} \\
\dot{f}_{D} &=& - \Gamma_{MD} f_{D} -\frac{1}{t_{D}} f_{D} + \frac{1}{t_{H_1}} f_{H_1} + \frac{1}{t_H} f_H \label{eq:req-infcyc-BB-D} \\
\dot{f}_{H} &=& \frac{p_{\rm{cycle}}}{t_{D}} f_{D} - \frac{1}{t_{H}} f_{H} \label{eq:req-infcyc-BB-C} \\
\dot{f}_S &=& \Gamma_{SI} f_I + \frac{1-p_{\rm{first}}}{t_{D_1}}f_{D_1} + \frac{1-p_{\rm{cycle}}}{t_{D}}f_{D} \label{eq:req-infcyc-BB-S},
\eea
where an overdot denotes a time derivative. The causal diamond measure instructs us to compare only the numbers of times various vacua are entered, rather than their volume fractions. Thus we need to compare the time integrals of the following ``incoming probability currents'' \cite{Linde:2006nw} (which can be read off directly from the rate equations as the positive sign terms on the right hand sides of the vacua of interest)
\bea
\dot{Q}_I &=& \Gamma_{IM} f_M \\
\dot{Q}_{H_1} &=& \frac{p_{\rm{first}}}{t_{D_1}}f_{D_1} \\
\dot{Q}_{H} &=& \frac{p_{\rm{cycle}}}{t_{D}}f_{D}.
\eea
In terms of co-moving volume, the entire multiverse starts out in $M$ and will eventually end up in $S.$ Thus the time integrals of the left hand sides of Eqs. (\ref{eq:req-infcyc-BB-I})-(\ref{eq:req-infcyc-BB-C}) from the initial time $t=0$ to $t=\infty$ are zero. The corresponding right hand sides then give us precisely the relations which we need to calculate the probabilities we are interested in. In particular, the probability to be in an emergent/cyclic pocket compared to the probability to be in a slow-roll inflationary one is given by
\bea
\frac{Q_H + Q_{H_1}}{Q_I} &=& \frac{\Gamma_{D_1M}}{\Gamma_{IM}}\frac{p_{\rm{first}}}{(1-p_{\rm{cycle}}+\Gamma_{MD} t_D)}\frac{1+\Gamma_{MD}t_D}{1+\Gamma_{MD_1}t_{D_1}} \\ &\approx& \frac{\Gamma_{D_1M}}{\Gamma_{IM}}\frac{p_{\rm{first}}}{(1-p_{\rm{cycle}}+\Gamma_{MD} t_D)}. \label{result}
\eea
Let us first discuss the case where the non-inflationary pocket corresponds to a non-cyclic, emergent-type universe. Then $p_{\rm{cycle}}=0$ and we obtain the following approximate relation
\be
\frac{Q_H + Q_{H_1}}{Q_I} \approx p_{\rm{first}} \frac{\Gamma_{D_1M}}{\Gamma_{IM}}. \label{resultapprox}
\ee
Given that the tunneling rate $\Gamma_{D_1M}$ is likely vastly larger than the up-tunneling rate $\Gamma_{IM},$ we confirm the expectation that it is much more likely to be in a 
non-inflationary pocket than in an inflationary one, as long as it is not exponentially suppressed to make it from the small-$H$ phase to the hot big bang phase. The same approximate formula will hold also for cyclic pockets where the probability $p_{\rm{cycle}}$ to make it from one cycle to the next is not very close to $1.$ However, as soon as the probability to cycle becomes close to one, the relative probability to be in a cyclic universe gets enhanced significantly \cite{Aguirre:2006ak,Johnson:2011aa}. This is because in that case a significant fraction of the co-moving volume of the multiverse experiences habitable conditions repeatedly inside each cyclic pocket. We can illustrate this effect by calculating the probability to be in the first cycle vs. a subsequent cycle,
\be
\frac{Q_{H_1}}{Q_H} = \frac{1-p_{\rm{cycle}}+\Gamma_{MD}t_D}{p_{\rm{cycle}}}\, .
\ee
Thus, when $p_{\rm{cycle}} \gtrsim \frac{1}{2},$ it is more likely to be in a later cycle rather than the first one after bubble nucleation.

We will add a few remarks that specifically concern the ekpyrotic/cyclic universe: the currently best understood incarnation of this model involves two scalar fields \cite{Lehners:2007ac}. The first scalar $\sigma$ drives the background dynamics, while the second scalar $s$ is responsible for generating scale-invariant perturbations. This second field is conjectured to have an {\it unstable} potential. This has two consequences: the first is that the probability to transition to the first hot big bang phase can only be large when the spread in field values of $s$ is small after tunneling, and the second is that, as discussed in \cite{Lehners:2008qe,Lehners:2009eg,Lehners:2011ig}, in this model typically $p_{\rm{cycle}} \ll 1.$ Let us first estimate $p_{\rm{first}}.$ We can adapt a very similar calculation performed by Garcia-Bellido {\it et al.} in \cite{GarciaBellido:1997te}, where they show that right after tunneling
\be 
\langle (\delta s)^2 \rangle^{1/2} \approx \frac{H_M}{2\pi\gamma^{1/2}},
\ee 
where, in the thin-wall limit, the parameter $\gamma$ is given by 
\be 
\gamma \approx \frac{2V_{,ssM}}{3H_M^2}+\frac{1}{8}H_M^2R_0^4(V_{,ssD1}-V_{,ssM}). 
\ee 
Here $V_{,ssM}$ and $V_{,ssD1}$ denote the effective masses of the $s$ field in the master and first dark energy vacua respectively, and $R_0$ is the size of the bubble at nucleation. It seems reasonable to assume that $V_{,ssM} \gg |V_{,ssD1}|$ since the master vacuum is very stable by definition. Moreover, since a bubble necessarily nucleates within a horizon-sized region, we have that $R_0 \lesssim H_F,$ so that one obtains
\be
\langle (\delta s)^2 \rangle^{1/2} \approx \frac{H_M^2}{\sqrt{V_{,ssM}}}.
\ee
Both $H_M$ being small and $V_{,ssM}$ being large help to make $\delta s$ small and thus $p_{\rm{first}}$ large. Thus, the two-field cyclic universe is highly preferred relative to inflationary pockets according to Eq. (\ref{resultapprox}). However, as mentioned above, the probability for a given co-moving volume to transition to a subsequent cycle is typically small in these models (even though the transitioning {\it physical} volume can be large due to the large net expansion that occurs during each cycle). Thus, the calculation above implies that it is most likely to find oneself in the very first cycle, rather than a later one. This is to some extent a shame, as it implies that certain of the attractive features of cyclic models (such as the ability to fine-tune parameters dynamically over many cycles \cite{Steinhardt:2006bf}) may be lost in this way. Of course, it is conceivable that the first cycle is still uninhabitable because certain parameters are not adjusted to the emergence of life yet, and that later cycles fine-tune themselves to allow for life. Under such circumstances, the earliest habitable cycle would be the most likely one. We should add that, although the instability discussed above as well as other potential instabilities related to the bounce appear as a drawback from the measure point of view, they have the potential of explaining certain cosmological quantities precisely by selecting out universes of a specific type -- see for instance \cite{Lehners:2010ug} for a possible explanation of the amplitude of the primordial density perturbations. In any case, though, the present discussion also motivates a closer re-examination of the single-field cyclic model, in which such an instability is absent. As is evident from Eq. (\ref{result}) with $p_{\rm{cycle}} = 1,$ if such models are indeed viable (for contrasting views see {\it e.g.} \cite{Tolley:2003nx} and \cite{Creminelli:2004jg}), they will vastly dominate over anything else in the landscape!

In concluding, we should discuss the main possible objection to the results presented above: it concerns the crucial aspect that underlies the preference of the multiverse for pocket universes that start out with a small expansion rate, namely that inside these pocket universes the null energy condition (NEC) must be violated in order to reach the hot big bang phase. This can happen either over an extended period of time, such as in galilean genesis or emergent scenarios \cite{Mulryne:2005ef,Lidsey:2006md}, or over a quick period, such as in the cyclic model. One may then wonder whether the NEC-violating aspect will turn out to be the Achilles heel of these models. At present, however, such a pessimistic view seems unwarranted: from an effective field theory point of view, there exist models, in particular the ghost condensate \cite{ArkaniHamed:2003uy} and galileons \cite{Nicolis:2008in}, that allow for NEC violation without leading to the appearance of ghosts. Even the inclusion of supersymmetry into these models does not lead to catastrophic instabilities \cite{Khoury:2010gb,Khoury:2011da}. Furthermore, string theory contains many objects that do violate the NEC, but are nevertheless well-behaved and stable -- examples include orientifolds and negative-tension orbifolds \cite{Polchinski:1996na,Lehners:2005su}. In fact, string theoretic models of inflation typically make use of these objects \cite{Cicoli:2011zz} (see also \cite{Steinhardt:2008nk}). In the braneworld incarnation of the cyclic model also, a negative-tension orbifold plane significantly affects the dynamics at the time of the bounce \cite{Khoury:2001bz,Lehners:2006pu}, while semi-classical calculations indicate that the bounce itself may be well-behaved \cite{Turok:2004gb}. Although this certainly does not constitute conclusive evidence yet that a bounce is possible in string theory, it certainly makes it conceivable. To the extent that the results of the present paper change the typically stated predictions of eternal inflation, they also motivate further work on the crucial quantum gravity issues of NEC violation and singularity resolution.

\begin{center}
\bf{Acknowledgements}
\end{center}

I would like to thank Matt Johnson for a careful reading of the manuscript and useful comments. I gratefully acknowledge the support of the European Research Council in the form of the Starting Grant numbered 256994.

\end{document}